\documentclass{myaa}

\usepackage[varg]{txfonts}

\usepackage{natbib}
\usepackage{graphicx}
\usepackage{epstopdf}
\usepackage{amssymb}
\usepackage{float}

\begin{document}

\title{Models of a protoplanetary disk forming in-situ the Galilean \\ and smaller nearby satellites before Jupiter is formed}

\author{Dimitris M. Christodoulou\inst{1,2}  
\and 
Demosthenes Kazanas\inst{3}
}

\institute{
Lowell Center for Space Science and Technology, University of Massachusetts Lowell, Lowell, MA, 01854, USA.\\
\and
Dept. of Mathematical Sciences, Univ. of Massachusetts Lowell, 
Lowell, MA, 01854, USA. \\ E-mail: dimitris\_christodoulou@uml.edu\\
\and
NASA/GSFC, Laboratory for High-Energy Astrophysics, Code 663, Greenbelt, MD 20771, USA. \\ E-mail: demos.kazanas@nasa.gov \\
}


\def\gsim{\mathrel{\raise.5ex\hbox{$>$}\mkern-14mu
                \lower0.6ex\hbox{$\sim$}}}

\def\lsim{\mathrel{\raise.3ex\hbox{$<$}\mkern-14mu
               \lower0.6ex\hbox{$\sim$}}}

\abstract{
We fit an isothermal oscillatory density model of Jupiter's protoplanetary disk to the present-day Galilean and other nearby satellites and we determine the radial scale length of the disk, the equation of state and the central density of the primordial gas, and the rotational state of the Jovian nebula. Although the radial density profile of Jupiter's disk was similar to that of the solar nebula, its rotational support against self-gravity was very low, a property that also guaranteed its long-term stability against self-gravity induced instabilities for millions of years.} 
\keywords{planets and satellites: dynamical evolution and stability---planets and satellites: formation---protoplanetary disks}

\authorrunning{ }
\titlerunning{Formation of Jupiter's Galilean and other nearby satellites}

\maketitle


\section{Introduction}\label{intro}

In previous work \citep{chr19a,chr19b}, we presented and discussed an isothermal model of the solar nebula capable of forming protoplanets long before the Sun was actually formed, very much as currently observed in high-resolution ($\sim$1-5~AU) observations of protostellar disks by the ALMA telescope \citep{alm15,and16,rua17,lee17,lee18,mac18,ave18,cla18,kep18,guz18,ise18,zha18,dul18,fav18,har18,hua18,per18,kud18,lon18,pin18,vdm19}.  Here we apply the same model to Jupiter's primordial disk that formed the Galilean satellites and probably a few additional neighboring moons. Our goal is to compare Jupiter's primordial nebula to the solar nebula and to find similarities and differences between the two disks that hosted gravitational potential minima in which the orbiting objects could form in relative safety over millions of years of evolution.

As was expected, the two model nebulae are different in their radial scale lengths and their sizes and central densities. Beyond these differences, the primordial disks are similar in the radial density and differential rotation profiles, albeit Jupiter's disk enjoys a lot lower rotational support against self-gravity than the solar nebula. The enhanced gas densities and the differential rotation speeds in Jupiter's compact disk signify that protosatellites could form around the protoplanet sooner than previously thought and long before Jupiter was formed.

In \S~\ref{theory}, we describe the analytic (intrinsic) and numerical (oscillatory) solutions of the isothermal Lane-Emden equation and the resulting model of the gaseous nebula. In \S~\ref{models2}, we apply the model nebula to the Galilean moons of Jupiter and a few other moons that may also have formed in the potential minima of the Jovian nebula. In \S~\ref{disc}, we summarize our results.

\section{Intrinsic and Oscillatory Solutions of the Isothermal Lane-Emden Equation with Rotation}\label{theory}

\subsection{Intrinsic Analytical Solutions}\label{imodel}


The isothermal Lane-Emden equation \citep{lan69,emd07} with rotation \citep{chr19a} takes the form of a second-order nonlinear inhomogeneous equation, viz.
\begin{equation}
\frac{1}{x} \frac{d}{dx} x \frac{d}{dx}\ln\tau \ + \ \tau \ = \ 
\frac{\beta_0^2}{2x}\frac{d}{dx}\left(x^2 f^2\right)\, ,
\label{main1}
\end{equation}
where $x\equiv R/R_0$, $\tau\equiv\rho /\rho_0$, $\rho(x)$ is the density and $\rho_0$ is the central density, $\beta_0\equiv \Omega_0 /\Omega_J$, $\Omega_J^2\equiv 2\pi G\rho_0$, $G$ is the gravitational constant, $f(x)$ is a function that describes the differential rotation of the disk, and the radial length scale $R_0$ is defined from the equation
\begin{equation}
R_0^2 \equiv \frac{c_0^2}{4\pi G\rho_0} = \frac{c_0^2}{2\Omega_J^2}\, ,
\label{length}
\end{equation}
where $c_0$ is the isothermal sound speed. The term $\Omega_J$ represents the gravitational (Jeans) frequency and 
the dimensionless rotation parameter $\beta_0$ measures centrifugal (i.e., rotational) support against self-gravity. 

Eq.~(\ref{main1}) has a large family of intrinsic analytical solutions that are independent of the imposed boundary conditions and take the form 
\begin{equation}
\tau (x) \ = \ \frac{\beta_0^2}{2}\cdot A x^{k-1} \ ,
\label{class2}
\end{equation}
and
\begin{equation}
f(x) \ = \ \frac{\sqrt{A\cdot g(x) + B}}{x} \ ,
\label{class3}
\end{equation}
where $A$, $B$, and $k$ are arbitrary integration constants and
\begin{equation}
g(x) \ \equiv \  \left\{ \begin{array}{cc} 
         x^{k+1}/(k+1) \, , & \ {\rm if} \ \ \ k \neq -1 \\
         \ln x \, , \ \ \ \ \ \ \ \ \ \ \ & \ {\rm if} \ \ \ k = -1 
         \end{array} \right. \ ,
\label{class4}
\end{equation}
implying that ~$dg/dx = x^k$ ~for all values of $k$.

Composite density models are built from equations (\ref{class2})-(\ref{class4}) by joining an inner ($x\leq x_1$) and an outer ($x\geq x_2$) flat-density region with an intermediate power-law region. These intrinsic solutions are shown by dashed lines in Figures \ref{fig1} and \ref{fig2} below. Continuity at the intersection points $x_1$ and $x_2 > x_1$ fixes the constants $A$ and $B$. The free parameters of the model then are the power-law index $k$, the rotation parameter $\beta_0$, and the two intersection points. There is however a very strong correlation (coefficient $r^2=0.99983$, $p$-value $= 2.466\times 10^{-3}$) between $\beta_0$ and $x_1$, viz.
\begin{equation}
\log_{10} x_1 = 1.1660 - 1.0384\cdot\log_{10}\beta_0 \ ,
\label{cor}
\end{equation}
that was also found in the modeling of our solar nebula and it allows us to drop $x_1$ from the list of free parameters. In the modeling of Jupiter's disk, we decided to keep $x_1$ as a free parameter and we used eq.~(\ref{cor}) to check that the correlation continues to hold in such a scaled-down version of a protoplanetary disk. We found that indeed the correlation holds true, so the parameters $x_1$ and $\beta_0$ below are strongly correlated.

\subsection{Numerical Oscillatory Solutions}\label{omodel}

When the Cauchy problem is solved numerically for eq.~(\ref{main1}) with the usual boundary conditions $\tau(0)=1$ and $[d\tau/dx](0)=1$ and with the same differential rotation profile as that of the analytic model, the Cauchy solutions cannot match the intrinsic solutions of \S~\ref{imodel}, precisely because they must satisfy the condition $\tau(0)=1$ which is incompatible with the intrinsic condition that $\tau(0)=\beta_0^2$ \citep{jea14}. Then the Cauchy solutions are attracted to the intrinsic solutions and oscillate permanently about these fundamental solutions. Such oscillatory density profiles are shown by solid lines in Figures \ref{fig1} and \ref{fig2} below. The numerical integrations were performed with the \textsc{Matlab} {\tt ode15s} routine \citep{sha97,sha99} and the optimization algorithm {\tt fminsearch} used the simplex search method of \cite{lag98} that does not implement any numerical or analytical gradients in its procedure. This optimization procedure is slow but also extremely stable numerically.

In the region of the inner core of the analytic models where the profiles are flat, the oscillatory solutions produce nearly equidistant density peaks. This is a generic feature of the Cauchy solutions. Such equidistant locations are also found in the inner three planets of our solar system. More importantly, the models of the solar nebula indicate that the two innermost protoplanets form in the inner core and the third protoplanet (Earth) forms just beyond point $x_1$, where the intrinsic density profile becomes a power law with index $k=-1.5$ \citep{chr19a}. These observations motivated us to look for similar features in Jupiter's disk model and to investigate an analogous arrangement of density peaks. It turns out that the small moon Thebe, Io, and Europa are in nearly equidistant orbits. So our {\it Model 1} of Jupiter's disk includes Thebe along with the Galilean satellites (Fig.~\ref{fig1}). 

The small inner moons of Jupiter are not in nearly equidistant orbits. Nevertheless, we also set out to optimize a model that includes the small moons Metis, Amalthea, and Thebe along with the Galilean satellites and the nearby moons Themisto and Himalia. This is despite reports that Amalthea has a cold water-ice composition, different than the other inner moons, and the hypothesis that it was not formed on its current orbit \citep{and05}. This {\it Model 2} (Fig.~\ref{fig2}) encounters trouble in fitting two of the inner smaller moons, likely because the inner orbits are not nearly equidistant. The more detailed Model 2 shares certain similarities with Model 1, but it shows some differences too, specifically in its radial scale length $R_0$ and the inner core's $x_1$ parameter (equivalently, the rotation parameter $\beta_0$). In both models, the Galilean satellites are fitted to consecutive density maxima to within deviations of $< 4$\%. 

Themisto and Himalia are in inclined orbits (45.8$^\circ$ and 30.5$^\circ$, respectively) relative to the rest of Jupiter's inner moons. Nevertheless, Model 1 shows that they can be fitted well to density maxima. For this reason, we included these neighboring moons in the optimization of Model 2. Their best-fit values show deviations of $< 0.7$\%, so these two moons do not affect the overall optimized result even if they are removed from the fit. We describe the detailed physical parameters of Models 1 and 2 in the next section.

\begin{figure}
\begin{center}
    \leavevmode
      \includegraphics[trim=0.2 0.2cm 0.2 0.2cm, clip, angle=0,width=10 cm]{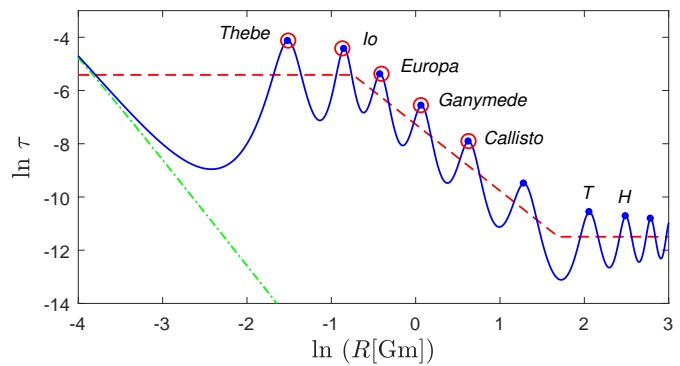}
      \caption{Equilibrium density profile for the midplane of Jupiter's primordial protoplanetary disk that formed the Galilean moons. The small inner moon Thebe was included in the fit. Themisto ($T$) and Himalia ($H$) are also shown in their peaks although they were not included in the fit. The best-fit parameters are $k=-1.5$, $\beta_0=0.0667$ (or, equivalently, $R_1=0.476$~Gm), and $R_2=5.42$~Gm. The radial scale length of the disk is $R_0=2061$~km. The Cauchy solution (solid line) has been fitted to the present-day moons of Jupiter so that its density maxima (dots) correspond to the observed semimajor axes of the orbits of the moons (open circles). The density maximum corresponding to the location of Ganymede was scaled to a distance of $R_G=1.0704$~Gm. The mean relative error of the fit is 1.6\%, affirming that this simple equilibrium model produces an incomparable match to the observed data points. The intrinsic solution (dashed line) and the nonrotating analytical solution (dash-dotted line) are also shown for reference. There exists an empty peak in the profile at $R = 3.609$~Gm ($\ln R = 1.2834$), where no moon is currently known.
\label{fig1}}
  \end{center}
\end{figure}

\section{Models of Jupiter's Protoplanetary Disk}\label{models2}

\subsection{Model 1}\label{model1}

In Fig.~\ref{fig1}, we show the optimized fit of Model 1. The inner small moon Thebe was used along with the Galilean satellites. We have effectively used only two free parameters ($k$ and $\beta_0$) to fit the current orbits of the five satellites, so it is not surprising that this fit is of very high quality (mean relative error of 1.6\%). Parameter $x_1$ is equivalent to $\beta_0$ and $x_2$ can be discarded from the list of free parameters because the density profile does not flatten out over the entire region of the Galilean satellites. These two parameters were retained only for comparison purposes with Model 2 below.

We find the following physical parameters from Model 1: $k=-1.5$, $\beta_0=0.0667$ (equivalently, $R_1=0.476$ Gm), and $R_2=5.42$ Gm, where $R_1$ and $R_2$ are the intersection points $x_1=231.07$ and $x_2=2632.5$, respectively, in physical units (1 Gm = $10^9$ m). The radial scale of the model was determined by fitting the density peak that corresponds to the orbit of Ganymede to its distance of 1.0704 Gm, and the scale length of the disk then is $R_0=2061$ km.

Model 1 is a scaled-down version of the model of the solar nebula \citep{chr19a} with the same power-law index ($k=-1.5$) but a much smaller scale length $R_0$. This is understood because the Jovian disk that extends out to the orbit of Callisto is not as extended as the solar nebula that reaches out to the orbits of Pluto and Eris. Furthermore, the model nebula is stable to self-gravitating instabilities because of its low value of $\beta_0=0.0667$ \citep[the critical value for the onset of nonaxisymmetric instabilities is $\beta_*\simeq 0.50$;][]{chr95}.

Outside the orbit of Callisto, there exist two neighboring moons, Themisto and Himalia. Although these moons were not used in the fit, it turns out that they are coincident with two peaks in the oscillatory density profile of Fig.~\ref{fig1} (denoted by $T$ and $H$). This leaves an empty spot at $R=3.609$ Gm, a density maximum in which a moon is not currently known to exist. We conclude that either the Jovian disk did not reach that far, or a moon failed to form at that location, or such a moon has not been discovered yet. This is an incentive for observers to search the region; after all, this region between Callisto ($R=1.9$ Gm) and Themisto ($R=7.5$ Gm) is too wide for it to be devoid of satellites.

\subsubsection{Physical Parameters of Model 1}\label{rhomax1}

Using the scale length of the disk ($R_0=2061$~km) 
in eq.~(\ref{length}), we can write the equation of state for the Jovian gas as
\begin{equation}
\frac{c_0^2}{\rho_0} \ = \ 4\pi G R_0^2 \ = \ 3.56\times 10^{10} 
{\rm ~cm}^5 {\rm ~g}^{-1} {\rm ~s}^{-2}\, ,
\label{crho1}
\end{equation}
where $c_0$ and $\rho_0$ are the local sound speed and the local density in the inner disk, respectively.
For an isothermal gas at temperature $T$, ~$c_0^2 = {\cal R} T/\overline{\mu}$, where $\overline{\mu}$ is the mean molecular weight and ${\cal R}$ is the
universal gas constant. Hence, eq.~(\ref{crho1}) can be rewritten as
\begin{equation}
\rho_0 \ = \ 2.33\times 10^{-3}\left(\frac{T}{\overline{\mu}}\right) \
{\rm ~g} {\rm ~cm}^{-3}\, ,
\label{trho1}
\end{equation}
where $T$ and $\overline{\mu}$ are measured in degrees Kelvin and 
${\rm ~g} {\rm ~mol}^{-1}$, respectively. 

For the coldest gas with $T \geq 10$~K 
and $\overline{\mu} = 2.34 {\rm ~g} {\rm ~mol}^{-1}$ (molecular hydrogen and
neutral helium with fractional abundances $X=0.70$ and $Y=0.28$ by
mass, respectively), we find that
\begin{equation}
\rho_0 \ \geq \ 0.01 \ {\rm ~g} {\rm ~cm}^{-3}\, .
\label{therho1}
\end{equation}
This high value implies that the conditions for protosatellite formation were already in place during the early isothermal phase \citep{toh02} of the Jovian nebula.

Using the above characteristic density $\rho_0$ of the inner disk
in the definition of ~$\Omega_J\equiv\sqrt{2\pi G\rho_0}$, we
can determine the Jeans frequency of the disk:
\begin{equation}
\Omega_J \ = \ 6.5\times 10^{-5} {\rm ~rad} {\rm ~s}^{-1}\, .
\label{thej1}
\end{equation}
Then, using the model's value $\beta_0 = 0.0667$ in the definition 
of ~$\beta_0\equiv \Omega_0 /\Omega_J$, we can determine the angular velocity of the uniformly-rotating core ($R_1\leq 0.476$~Gm), viz.
\begin{equation}
\Omega_0 \ = \ 4.3\times 10^{-6} {\rm ~rad} {\rm ~s}^{-1}\, .
\label{theom1}
\end{equation}
For reference, this value of $\Omega_0$ corresponds to an orbital period 
of 16.85~d. This value is close to the present-day orbital period of Callisto (16.69~d). Thus, the core of the Jovian nebula was rotating about as slowly as Callisto is presently revolving around Jupiter.

\begin{figure}
\begin{center}
    \leavevmode
      \includegraphics[trim=0.2 0.2cm 0.2 0.2cm, clip, angle=0,width=10 cm]{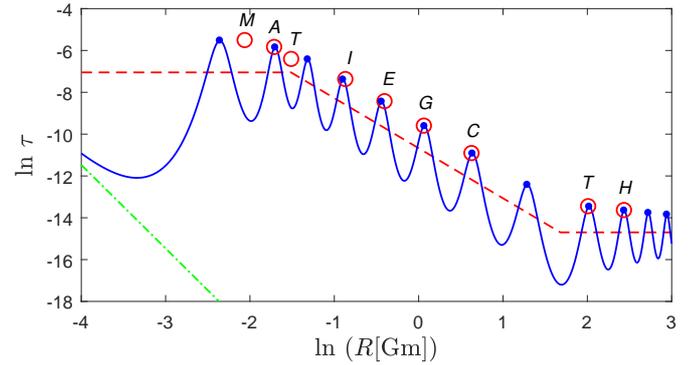}
      \caption{Equilibrium density profile for the midplane of Jupiter's primordial protoplanetary disk that formed the Galilean moons. The small inner moons Metis, Amalthea, and Thebe, as well as Themisto and Himalia were also included in this fit. The best-fit parameters are $k=-1.4$, $\beta_0=0.0295$ (or, equivalently, $R_1=0.220$~Gm), and $R_2=5.37$~Gm. The radial scale length of the disk is only $R_0=368$~km. The Cauchy solution (solid line) has been fitted to the present-day moons of Jupiter so that its density maxima (dots) correspond to the observed semimajor axes of the orbits of the moons (open circles). The density maximum corresponding to the location of Ganymede was scaled to a distance of $R_G=1.0704$~Gm. The mean relative error of the fit is 13\%, all of which comes from the errors in the positions of Metis and Thebe. The remaining moons have positional errors $< 4$\%. The intrinsic solution (dashed line) and the nonrotating analytical solution (dash-dotted line) are also shown for reference. There exists an empty peak in the profile at $R = 3.625$~Gm ($\ln R = 1.2878$), where no moon is currently known. 
\label{fig2}}
  \end{center}
\end{figure}

\subsection{Model 2}\label{model2}

In Fig.~\ref{fig2}, we show the optimized fit of Model 2 that includes three inner moons (Metis\footnote{Another minor moon, Adrastea, orbits inside the potential minimum of Metis and it is omitted from the fit.}, Amalthea, and Thebe), the Galilean satellites, and the neighboring moons Themisto and Himalia. We have used four free parameters ($k$, $\beta_0$, $x_1$, and $x_2$) to fit the current orbits of the 9 satellites, and the fit is very good for 7 of them. The model has difficulty fitting the orbits of Metis and Thebe (errors $\sim$20\%). The mean relative error is 13\%, but all of it comes from the orbits of Metis and Thebe. The remaining 7 moons are fitted to much better than 4\% deviations each. Parameter $x_1=597.46$ can be discarded from the list of free parameters because it is strongly correlated to $\beta_0=0.0295$ (see eq.~(\ref{cor})).

We find the following physical parameters from Model 2: $k=-1.4$, $\beta_0=0.0295$ (equivalently, $R_1=0.220$ Gm), and $R_2=5.37$ Gm (scaled from $x_2=14562$). The radial scale of the model was determined by fitting the density peak that corresponds to the orbit of Ganymede to its distance of 1.0704 Gm, and the scale length of the disk then is $R_0=368$ km.

Model 2 is a scaled-down version of the model of the solar nebula \citep[$k=-1.5$;][]{chr19a} with approximately the same power-law index ($k=-1.4$) but a much smaller scale length $R_0$. This is understood because this Jovian disk extends out to the orbit of Himalia and it is much smaller than the size of the solar nebula. Furthermore, this model nebula is stable to self-gravitating instabilities because of its lower value of $\beta_0$ as compared to Model 1.

The orbits of Themisto and Himalia were used in the fit of Model 2. But Model 2 still predicts an empty density maximum very close to the location of another empty maximum shown in Model 1. This empty spot is located at $R=3.625$ Gm. Once again, this is an incentive for observers to search the wide empty region around Jupiter between $R=1.9$ Gm (Callisto) and $R=7.5$ Gm (Themisto) for an additional satellite of any size.

\subsubsection{Physical Parameters of Model 2}\label{rhomax2}

Using the scale length of the disk ($R_0=368$~km) 
in eq.~(\ref{length}), we can write the equation of state for the Jovian gas as
\begin{equation}
\frac{c_0^2}{\rho_0} \ = \ 4\pi G R_0^2 \ = \ 1.14\times 10^{9} 
{\rm ~cm}^5 {\rm ~g}^{-1} {\rm ~s}^{-2}\, .
\label{crho2}
\end{equation}
For an isothermal gas with mean molecular weight $\overline{\mu}$ at temperature $T$, eq.~(\ref{crho2}) can be rewritten as
\begin{equation}
\rho_0 \ = \ 7.30\times 10^{-2}\left(\frac{T}{\overline{\mu}}\right) \
{\rm ~g} {\rm ~cm}^{-3}\, ,
\label{trho2}
\end{equation}
where $T$ and $\overline{\mu}$ are measured in degrees Kelvin and 
${\rm ~g} {\rm ~mol}^{-1}$, respectively. 

For the coldest gas with $T \geq 10$~K 
and $\overline{\mu} = 2.34 {\rm ~g} {\rm ~mol}^{-1}$ (as in \S~\ref{rhomax1}), we find that
\begin{equation}
\rho_0 \ \geq \ 0.31 \ {\rm ~g} {\rm ~cm}^{-3}\, .
\label{therho2}
\end{equation}
This high value again implies that the conditions for protosatellite formation were already in place during the early isothermal phase \citep{toh02} of the Jovian nebula.

Using the above characteristic density $\rho_0$ of the inner disk
in the definition of ~$\Omega_J\equiv\sqrt{2\pi G\rho_0}$, we
can determine the Jeans frequency of the disk:
\begin{equation}
\Omega_J \ = \ 3.6\times 10^{-4} {\rm ~rad} {\rm ~s}^{-1}\, .
\label{thej2}
\end{equation}
Then, using the model's value $\beta_0 = 0.0295$ in the definition 
of ~$\beta_0\equiv \Omega_0 /\Omega_J$, we can determine the angular velocity of the uniformly-rotating core ($R_1\leq 0.220$~Gm), viz.
\begin{equation}
\Omega_0 \ = \ 1.1\times 10^{-5} {\rm ~rad} {\rm ~s}^{-1}\, .
\label{theom2}
\end{equation}
For reference, this value of $\Omega_0$ corresponds to an orbital period 
of 6.819~d. This value is close to the present-day orbital period of Ganymede (7.155~d). Thus, in this model, the core of the Jovian nebula was rotating about as slowly as Ganymede is presently revolving around Jupiter.

\begin{table*}
\caption{Comparison of Models 1 and 2 of Jupiter's protoplanetary disk}
\label{table1}
\begin{tabular}{llll}
\hline
Property & Property & Jupiter's & Jupiter's \\
Name     & Symbol (Unit) & Model 2 & Model 1 \\
\hline
Density power-law index & $k$  &   $-1.4$  	     & $-1.5$    \\
Rotational parameter & $\beta_0$  &    0.0295 	       &  0.0667   \\
Inner core radius & $R_1$ (Gm)   &   0.220  	       &  0.476    \\
Outer flat-density radius & $R_2$ (Gm) &   5.37        	   &  5.42   \\
Scale length & $R_0$ (km)    &   368        	   &  2061   \\
Equation of state & $c_0^2/\rho_0$ (${\rm cm}^5 {\rm ~g}^{-1} {\rm ~s}^{-2}$) & $1.14\times 10^9$ & $3.56\times 10^{10}$    \\
Minimum core density for $T=10$~K, $\overline{\mu} = 2.34$ & $\rho_0$ (g~cm$^{-3}$)         &    0.31   			&  0.01   \\
Isothermal sound speed for $T=10$~K, $\overline{\mu} = 2.34$ & $c_0$ (m~s$^{-1}$) & 188 & 188 \\
Jeans gravitational frequency & $\Omega_J$ (rad~s$^{-1}$)    &    $3.6\times 10^{-4}$ & $6.5\times 10^{-5}$    \\
Core angular velocity & $\Omega_0$ (rad~s$^{-1}$)    &    $1.1\times 10^{-5}$ 	& $4.3\times 10^{-6}$    \\
Core rotation period & $P_0$ (d)                                 &    6.819 	   			&  16.85   \\
Maximum disk size & $R_{\rm max}$ (Gm)                &    12 	   			&   12  \\
\hline
\end{tabular}
\end{table*} 

\subsubsection{Comparison of Models 1 and 2}\label{comp}

The agreement between Model 1 and Model 2 is satisfactory but not precise (Table~\ref{table1}). In both Models 1 and 2, the power-law density profile appears to have an index of $k\approx -1.5$ (also similar to that of the solar nebula) and the outer flat-density radius appears to be $R_2\approx 5.4$ Gm. The small inner moons appear to have formed within the inner flat core, whereas the Galilean satellites all lie along the gradient of the density profile of the Jovian nebula; and Themisto and Himalia fall within the outer flat-density region (assuming that the primordial disk extended out that far).

There are however some notable differences in the remaining physical parameters of the two models: 
\begin{itemize}
\item[(a)] Centrifugal (rotational) support is about half in Model 2 than in Model 1. Both values of $\beta_0$ are very low, so these model nebulae are in no danger of suffering nonaxisymmetric instabilities over their entire lifetimes. 
\item[(b)] The size of the inner core in Model 2 is about half of that in Model 1 in physical units (Gm). This difference reflects the widely different scale lengths of the two models ($R_0=2061$ km versus $R_0=368$ km, in Model 1 and 2, respectively). The smaller core of Model 2 can easily accommodate the small inner moons (Fig.~\ref{fig2}); and then there is even more space for them in the core of Model 1 in which we considered only the small moon Thebe (Fig.~\ref{fig1}). However, the smaller inner moons are not arranged in approximately equidistant orbits and the models have trouble fitting their locations well.
\item[(c)] As a result of the above differences, the rotation period of the model nebula appears to be somewhere between the present-day orbital period of Ganymede and Callisto (7.16 d and 16.7 d, respectively). In any case, the rotation of Jupiter's disk was slow. It is interesting that the more detailed Model 2 predicts a core rotation period close to the orbital period of the largest satellite (Ganymede); the same coincidence also occurs in our model of the solar nebula \citep{chr19a}.
\item[(d)] The central density of the gas and the equation of state appear to be different by one order of magnitude between the two models (Table~\ref{table1}). In either case, the local densities in the compact Jovian nebula were significantly higher than those in the solar nebula, and this property indicates that protosatellites could form very early in the evolution of the Jovian system and certainly long before Jupiter itself. 
\end{itemize}
The models support a  ``bottom-up'' hierarchical formation in which protosatellites are seeded early inside protoplanetary disks and long before their protoplanets are fully formed; and then, these compact systems complete their formation in $< 0.1$ Myr and long before the protosun becomes fully formed. There is already some support for this scenario from the observation of an asymptotically flat rotation curve (as in all models, including that of the solar nebula) in the Class 0 very young protostellar system HH 211-mms in Perseus \citep{lee18}; and from the observations of extremely young protostellar systems conducted by \cite{gre10} and recently of the system TMC1A, in particular, whose age is estimated to be $< 0.1$ Myr \citep{har18}.

\section{Summary}\label{disc}

We have constructed isothermal differentially-rotating protoplanetary models of the Jovian nebula, the primordial disk in which the Galilean satellites and probably some of their neighboring smaller moons were formed. We first analyzed Model 1 of the Galilean satellites along with the small inner moon Thebe that is roughly equidistant with Io and Europa. This baseline model indicates that Jupiter's protoplanetary disk had very little centrifugal support against self-gravity, which signifies that the disk was stable against dynamical non-axisymmetric instabilities. Thus, an obvious and strong constraint of the formation of these moons is clearly satisfied.

In Model 2, we added all major moons out to Himalia and we attempted another optimization of the physical properties of this Jovian nebula. Some parameters remain approximately the same (the power-law index $k\approx -1.5$ and the radius of the outer flat-density region $R_2\approx 5.4$ Gm). The remaining parameters change, but not in an alarming way: when 9 moons were considered in Model 2, the radial scale length $R_0$ decreased by a factor of $\sim$5.6, the inner core radius $R_1$ and the rotation parameter $\beta_0$ of the nebula were approximately halved, and the central density $\rho_0$ increased by an order of magnitude. In both models, however, Jupiter's primordial disk appears to be stable and long-lived. Thus, the moons can form early in the evolution of the nebula and long before Jupiter manages to pull its gaseous envelope on to its solid core.

Comparison between these models (Table~\ref{table1}) supports the following structural properties of the Jovian system:
\begin{itemize}
\item[1.] Protosatellites form at the bottom of a ``bottom-up'' scenario before their protoplanets are fully formed. Then these protoplanetary systems complete their formation before the protosun is actually formed at the center of the solar nebula.
\item[2.] The small inner moons of Jupiter are not arranged in a regular pattern of approximately equidistant orbits. As a result, our models have difficulties in fitting their locations. The orbits of these small moons may have been perturbed after the primordial nebula had dissipated away.
\item[3.] The current orbital period of the largest orbiting body (Ganymede for the Jovian system and Jupiter for the solar system) may be important in that it reveals the rotation period of the core of the corresponding primordial nebula.
\item[4.] The moons Themisto and Himalia neighboring the Galilean satellites are not part of the Galilean group. In both models, the Galilean satellites form along the density gradient of the intrinsic solution of the Jovian nebula, whereas these two smaller moons form in the outer flat region of the nebular density profile.
\item[5.] In the oscillatory density profile of Jupiter's model nebula, there is an empty peak at a radius of about 3.6 Gm where no moon is currently known. This peak falls in the wide region between Callisto and Themisto (1.9-7.5 Gm) and it is actually surprising that no moons have been discovered in such a vast area. If a moon is ever found at about this distance, it will definitely be another member of the Galilean satellites.
\end{itemize}

\end{document}